\documentstyle{article}
\title{\bf Extreme Cosmic String}
\author{ Pedro F. Gonz\'alez-D\'{\i}az.\\
Centro de F\'{\i}sica "Miguel Catal\'an",\\
Instituto de Matem\'aticas y F\'{\i}sica Fundamental\\
Consejo Superior de Investigaciones Cientificas\\
Serrano 121, 28006 Madrid (SPAIN)\\
}
\date{}
\begin{document}
\maketitle
\large
\setlength{\baselineskip}{0.5cm}

\begin{center}
{\bf Abstract}
\end{center}

\noindent This paper deals with the geometry of supermassive cosmic
strings.
We have used an approach that enforces the
spacetime of cosmic strings to also satisfy the
consevation laws of a
cylindric gravitational topological defect, that is a spacetime
kink. In the simplest case of kink number unity the entire
energy range of supermassive strings becomes then quantized so
that only cylindrical defects with linear energy density
$G\mu=\frac{1}{4}$
(critical string) and $G\mu=\frac{1}{2}$ (extreme
string) are allowed to occur
in this range. It has been seen that the internal spherical
coordinate $\theta$ of the string metric embedded in an
Euclidean three-space
also evolves on imaginary
values, leading to the creation of a covering shell of broken phase
that protects the core with trapped energy, even for $G\mu=
\frac{1}{2}$. Then, the conical singularity becomes a
removable horizon singularity. We re-express the extreme
string metric in the Finkelstein-McCollum standard form and
remove the geodesic incompleteness by using the Kruskal
technique. The $z$=const sections of the resulting metric
are the same as the hemispherical section of the metric of a
De Sitter kink. Some physical consequences from these
results, including the possibility that the extreme string
drives inflation and thermal effects in its core,
are also discussed.

\pagebreak

\section{Introduction}

Topological defects consisting of confined regions of false vacuum
can occur in gauge theories
with spontaneous symmetry breaking [1].\\
Among such defects, cosmic
strings trapped during the breaking of a local $U(1)$ gauge symmetry
are of particular interest [1,2]. If local strings appeared in
phase transitions in the early universe, they could
have served as seeds for the formation of galaxies and other
larger scale structures we now are able to observe [3]. Recently,
the possibility that inflation can be driven in the core of
topological defects has also been advanced [4,5]. In the case of
local cosmic strings this extremely interesting possibility will
critically depend [5] on the strength of their gravity
coupling $G\mu$, where $\mu$ is the string mass per unit length.

Static defect solutions occur in simple models of the form
\[V(\varphi)=\frac{1}{4}\lambda(\varphi_{a}\varphi_{a}-\eta^{2})^{2},
a=1,...,N.,\]
where $\eta$ is the symmetry breaking scale, and $N=2$ for cosmic strings.
Equalizing [5] the core radius $\delta_{0}
\sim\eta V_{0}^{-\frac{1}{2}}$,
where $V_{0}=\frac{1}{4}\lambda\eta^{4}$,
and the horizon size corresponding
to the vacuum energy $V_{0}$,
$H_{0}^{-1}=(\frac{3}{8\pi GV_{0}})^{\frac{1}{2}}$,
one obtains using $\mu\sim\eta^{2}$ a
value for the string mass per unit length $\mu_{I}\sim
\frac{1}{G}$, such that for $\mu>\mu_{I}$ inflation could be
generated in the core.

Nevertheless, the concepts of radius and
mass per unit length for a source
like the string core are not unambiguously defined [6],
and one would expect that even for the extreme case
where $\mu=\mu_{e}=\frac{1}{2G}$
inflation could not be ensured to be driven in the
core of the cosmic string.
On the other hand, an extreme
supermassive string with $2G\mu_{e}=1$ does not seem to
exist because it would correspond to a situation where all
the exterior broken phase is collapsed into the core, leaving
a pure false-vacuum phase in which the picture of a cosmic
string with a core region of trapped energy is lost [6-9].
This happens in all considered string metrics, i.e. for the
Hiscock-Gott metric [10,11] and for the Laguna-Garfinkle [8]
and Ortiz [9] metrics, in all the cases the spacetime possesses
an unwanted singularity which cannot be smoothed out.

In this paper we enforce the interior string metric to describe
a cylindrically-symmetric
gravitational kink, i.e. an allowed gravitational topological defect
which can move about spacetime but cannot be removed without cutting [12],
and is characterized by a conserved kink number measuring the
number of times the light cone tips over on a boundary.
We show that in this case  the picture of a
cosmic string with a core region of trapped energy is still
retained even at the extreme value $2G\mu=1$, and that the
conical singularity becomes then the apparent singularity
(event horizon) of a De Sitter kink. This horizon singularity
can be removed by a suitable Kruskal extension. The
resulting extreme supermassive string would then be able to drive
a gravitational inflationary process,
without any fine tuning of the initial conditions.

The paper is outlined as follows. In section 1 we construct
an explicit metric for the supermassive cosmic string kink,
and discuss the constraints that the existence of the kink
imposes on the internal geometry of the string. As a result
of conservation of kink number,
the energy of the supermassive cosmic string becomes quantized
so that it can only take on values $4G\mu=1$ and
$2G\mu=1$. We then re-express the metric of the cosmic
string kink in standard form and obtain an analytical expression
for the relevant time parameter entering that metric in section
3. The geodesic incompleteness of the
standard metric is removed in section 4 by maximally-extending
this metric using the Kruskal technique. In section 4 we also
show that the $z=$const. sections of the
geodesically complete metric of a $2G\mu=1$
cosmic-string kink describes a hemispherical section of a kinky De
Sitter spacetime which may allow an eternal inflationary process [13],
and study the quantum creation of particles in
the one-kink extreme cosmic
string. Throughout the paper we use units
so that $\hbar=c=1$.

\section{The Spacetime of a Cosmic String Kink}
\subsection{Static Metric}
\setcounter{equation}{0}

The static, cylindrically symmetric internal metric of a straight
cosmic string, which is an exact solution of Einstein equation,
is given by [10,11]
\begin{equation}
ds^{2}=-d\tau^{2}+d\rho^{2}+dz^{2}+r_{*}^{2}\sin^{2}\frac{\rho}{r_{*}}d\phi^{2},
\end{equation}
with $-\infty<\tau<\infty$, $-\infty<z<\infty$, $0\leq\phi<2\pi$,
$0\leq\rho\leq r_{*}\arccos(1-4G\mu)$, and
\begin{equation}
r_{*}=(8\pi G\epsilon)^{-\frac{1}{2}},
\end{equation}
where $\epsilon$ is the uniform string density, out to some
cylindrical radius $\rho_{0}$.

Both the interior metric (2.1) and the exterior metric [10,11],
\begin{equation}
ds^{2}=-d\tau^{2}+d\rho'^{2}+dz^{2}+(1-4G\mu)^{2}\rho'^{2}d\phi^{2},
\end{equation}
define two-surfaces at $z=$const, $\tau=$const, which can be
simultaneously visualized by embedding the metrics in an Euclidean
three-space [11]. Then, the geometries of such surfaces are,
respectively, that of a spherical cap (interior region) and that
of a cone with deficit angle $\Delta=8\pi G\mu$ in the exterior
vacuum region. The interior spherical cap becomes a
hemisphere for $G\mu=\frac{1}{4}$ and a sphere for
$G\mu=\frac{1}{2}$. In what follows, these embeddings
will be denoted as embedded hemispherical or spherical
geometries.

We shall write now the internal metric (2.1) in a form which
is best suited for performing a kink extension.
Thus, the change of coordinate
\begin{equation}
r=r_{*}\sin\frac{\rho}{r_{*}}
\end{equation}
shows more transparently the invariance of the line element (2.1)
under rotation about $\phi$ and traslation along $z$ of the
underlying cylindrical symmetry (i.e. under the two Killing
vectors $\xi=\partial_{\phi}$ and $\zeta=\partial_{z}$ [14]),
and gives rise to a singularity at $r=r_{*}$; i.e.:
\begin{equation}
ds^{2}=-d\tau^{2}+\frac{dr^{2}}{1-\frac{r^{2}}{r_{*}^{2}}}+dz^{2}+r^{2}d\phi^{2}.
\end{equation}

The divergence that appears in
(2.5) as $r$ tends to $r_{*}$ would correspond to
an apparent singularity and should therefore be removable by an
appropriate coordinate transformation. It follows from (2.2)
and (2.4) that values of coordinate $r$
larger than $r_{*}$ should be associated with purely
imaginary values of $r_{*}$ and, therefore with negative string
densities, $\epsilon<0$. It will be seen in section 3 that,
if the internal spacetime of a cosmic string satisfies the
conservation law of a kink, then
such imaginary values of $r_{*}$ would imply Euclidean
continuation of time $\tau$ so that it becomes generally complex,
and hence instantonic "tunneling" into the exterior
broken-symmetry phase. One would regard the kink extension
of metric (2.5) as a way to determine the precise
extent of such a continuation.

A coordinate change which
still transparently shows invariance under the Killing vectors
and leads to no apparent interior singularity is
\begin{equation}
u=\frac{\tau}{r_{*}}+\arcsin\frac{r}{r_{*}},
v=\frac{\tau}{r_{*}}-\arcsin\frac{r}{r_{*}}.
\end{equation}
In terms of coordinates $u$ and $v$ we obtain for the interior metric
\begin{equation}
ds^{2}=-r_{*}^{2}dudv+dz^{2}+r^{2}d\phi^{2},
\end{equation}
where
\begin{equation}
r=r_{*}\sin[\frac{1}{2}(u-v)].
\end{equation}
In any case, the exterior metric still keeps a conical singularity.

\subsection{Kink Nonstatic Extension}

The regular metric (2.7) would trivially describe the maximally
extended interior region of a cosmic string.
The structure of the string core is then determined by
the matter content of the gauge theory
(e.g the Abelian Higgs model of Nielsen
and Olesen [15]) and the coupling $G\mu$,
and should only lodge
real energy from the unbroken gauge phase. Nevertheless,
metric (2.5) can be analytically extended into some limited
region with the broken phase if we make this metric allow
the lightcones to tip over on the boundaries so that one
conserved kinky topological defect describable by
relativistic metric twists is present [12].

The relationship between a possible cosmic string kink and
the apparent singularity in metric (2.5) can be appreciated
by considering [16] a cylindrically-symmetric spacetime with
manifest invariance under the action of the Killing vectors
$\xi$ and $\zeta$ of the form
\begin{equation}
ds_{\alpha}^{2}=\cos 2\alpha(-dt^{2}+dr^{2})-2\sin 2\alpha
dtdr+dz^{2}+r^{2}d\phi^{2},
\end{equation}
where $\alpha$ denotes the angle of tilt of the spacetime
lightcones. We shall then ensure the presence of one kink
(i.e. a gravitational topological defect with kink number
unity) by requiring $\alpha$ to monotonously vary from
$0$ to $\pi$, starting from $\alpha(0)=0$.

Metric (2.9) is nonstatic (as it is not invariant under time
reversal $t\rightarrow-t$) and
can be transformed into the static metric (2.5) if we set
\begin{equation}
\sin\alpha=\frac{r}{2^{\frac{1}{2}}r_{*}},
\end{equation}
and change the time coordinate so that $\tau=t+G(\omega)$, where
\begin{equation}
d\omega=d\tau+F(r)dr,
\end{equation}
with $F(r)$ a given function of $r$. Denoting $\frac{dG(\omega)}{d\omega}$
by $G'$ it follows
\begin{equation}
dt=d\tau(1-G')-G'F(r)dr.
\end{equation}
Metric (2.5) can then be obtained from (2.9) if
\begin{equation}
G'=1-\frac{k_{1}}{\cos^{\frac{1}{2}}2\alpha}, k_{1}=\pm 1
\end{equation}
\begin{equation}
G'F(r)=\tan 2\alpha.
\end{equation}
Note that both $F(r)$ and $G'$ are singular at $r=r_{*}$. This still
reflects the singular character of metric (2.5). Moreover,
since this metric can be transformed into metric (2.9) by (2.10)-(2.14),
the latter metric
becomes also a solution of the same Einstein equation
as for (2.5), expressed in terms of the new time coordinate $t$.

\subsection{The Broken-Phase Shell}

Let us consider the
limitations that the functional form (2.10) imposes on the
internal geometry of a supermassive cosmic string. From (2.4) and
(2.10) it is obtained
\begin{equation}
\cos^{2}\theta = \cos 2\alpha,
\end{equation}
where we have denoted [11] $\theta=\frac{\rho}{r_{*}}$, so that
the $\tau$,$z$ = const. sections of metric (2.1) becomes
$r_{*}^{2}(d\theta^{2}+\sin^{2}\theta d\phi^{2})$ for some
$\theta$-interval, and $\alpha$
monotonously varies from $0$ to $\pi$.

The question now is,
how does $\theta$ vary under the complete one-kink
variation of $\alpha$ from $0$
to $\pi$?. From (2.15) it follows that monotonous variation of
$\alpha$ from $0$ to $\frac{\pi}{4}$, and from $\frac{3\pi}{4}$
to $\pi$ induces monotonous variation of $\theta$, respectively,
from $0$ to $\frac{\pi}{2}$ and from $\frac{\pi}{2}$ to $\pi$
(or, likewise, from $\pi$ to $\frac{\pi}{2}$ and from $\frac{\pi}{2}$
to $0$), i.e. the entire two-sphere.
Note that the corresponding
induced variations of $\theta$ in the interval
($\pi$,$2\pi$) are also allowed but, since $\theta$ appears in the
metric in the form $\sin^{2}\theta$, these variations would lead to
the same geometrical situations as from variations of $\theta$
in the interval (0,$\pi$).
Variation of $\alpha$ from $\frac{\pi}{4}$ to $\frac{3\pi}{4}$
induces variation of $\theta$ only along its imaginary axis,
first from $\frac{\pi}{2}$ to $\frac{\pi}{2}+i\ln(2^{\frac{1}{2}}\pm 1)$
(at $\alpha=\frac{\pi}{2}$), and then to $\frac{\pi}{2}$ again.
The choice of sign in the argument of the $\ln$ at the extremum
value of $\theta$ corresponding to $\alpha=\frac{\pi}{2}$ should
be made as follows. On the interval where $\theta$ is complex, i.e.
$\theta=\frac{\pi}{2}+i\theta_{i}$,
with $\ln(\sqrt{2}-1)\leq\theta_{i}\leq\ln(\sqrt{2}+1)$,
we have (see Refs. [10,11])
\begin{equation}
G\mu=\frac{1}{4}+\frac{i}{4}\sinh\theta_{i}
\end{equation}
\begin{equation}
r=ir_{*}\coth\theta_{i}
\end{equation}
Taking into account (2.17) it follows that the imaginary part of
(2.16) corresponds to some real mass per unit imaginary length, or
equivalently, to some tachyonic mass per real unit length. Thus, once
the critical string mass $G\mu=\frac{1}{4}$ is reached, the interior of
the kinky string starts developing a new real region
up to a maximum width
$(2^{\frac{1}{2}}-1)r_{*}$, covering the entire
embedded hemisphere of the $\tau$,$z$=const. sections, where
the gauge symmetry is broken at the given scale $\varphi=\pm\eta$. Then,
the symmetry $\eta\rightarrow-\eta$ of the broken phase will
imply an identification $\frac{\pi}{2}+i\ln(2^{\frac{1}{2}}+1)$
$\rightarrow\frac{\pi}{2}+i\ln(2^{\frac{1}{2}}-1)$ on the extremum
value of $\theta_{i}$ that corresponds to $\alpha=\frac{\pi}{2}$.
We can therefore choose complex $\theta$ to vary first from
$\frac{\pi}{2}$ to $\frac{\pi}{2}+i\kappa\ln(2^{\frac{1}{2}}+1)$
as $\alpha$ goes from $\frac{\pi}{4}$ to $\frac{\pi}{2}$, and
then from $\frac{\pi}{2}+i\kappa\ln(2^{\frac{1}{2}}-1)$ to
$\frac{\pi}{2}$ again as $\alpha$ goes from $\frac{\pi}{2}$ to
$\frac{3\pi}{4}$, either for $\kappa=\pm 1$. This strip
corresponds to the
maximum analytical extension beyond $r=r_{*}$ of the real region
described by the internal string metric (2.5) which is
compatible with the presence of one kink.

Now, the maximum value of the internal spherical coordinate
$\theta$, $\theta_{M}$, is related to the string mass per
unit length $\mu$ by [10,11]
\[\theta_{M}=\arccos(1-4G\mu).\]
Then, in order to ensure the occurrence of one kink in the
cosmic string interior, we must have $\theta_{M}=\pi$ and hence
$G\mu=\frac{1}{2}$; or in other words, the presence of one kink
implies a {\it quantization} of the supermassive cosmic
string so that only the critical string at
$G\mu=\frac{1}{4}$ with internal embedded hemispherical
geometry (no kink present), and the extreme
string at $G\mu=\frac{1}{2}$ with internal embedded spherical
geometry (one kink present)
are allowed to exist along the entire interval,
$\frac{\pi}{2}\leq\theta\leq\pi$, of possible classical
supermassive cosmic strings.

If we ensure the presence of one kink, then no string exterior
(except the portion which is incorporated into the interior
metric by the complex evolution of $\theta$) is possible or
needed [10,11], and the conical singularity becomes a removable
horizon singularity on the core surface at $r=r_{*}$,
separating the two gauge phases that make up the extended
string interior. Thus, all of the possible geometry of
the extreme string can be regarded as describable by a
spacetime which is ${\bf R}^{2}\times S^{2}$, showing still
the picture of a cosmic string with a core region
of trapped energy surrounded by a shell of true vacuum
protecting the string from dissolving in the background
phase with unbroken symmetry.

\section{The Lightcone Configuration}
\setcounter{equation}{0}

The results of section 2 allow us to deal with the interior metric
of an extreme supermassive cosmic string kink, $2G\mu=1$,
in a similar fashion to as it is done in the cases
of the black hole kink [16] or the De Sitter kink [17].

The interior spacetime metric of one such
extreme string has still a geodesic incompleteness
at $r=r_{*}$, and can only be described by using a number of
different coordinate patches. Since $\sin\alpha$ is given
by Eq. (2.10) and cannot exceed unity, it follows that
$0\leq r\leq A\equiv\sqrt{2}r_{*}$, so that $\alpha$ varies
only from 0 to $\frac{\pi}{2}$. In order to have an entire
one-kink gravitational defect, we need then a second coordinate
patch which would describe the other half of the
$\alpha$ interval, $\frac{\pi}{2}\leq\alpha\leq\pi$.
The identification  and
distinction of such two patches can be achieved by transforming
(2.9) into the standard metrical form proposed by
Finkelstein and McCollum [16], adapted here to cylindric symmetry.
To accomplish such a transformation, it is convenient to
introduce a new time coordinate, such that
\begin{equation}
\bar{t}=t+G(\sigma)
\end{equation}
\begin{equation}
\cos 2\alpha(1-G^{;2}F_{\sigma}^{2}(r))+2\sin 2\alpha G^{;}F_{\sigma}(r)=0,
\end{equation}
where the new variable $\sigma$ and the new functionals $F_{\sigma}(r)$
and $G\equiv G(\sigma)$ are defined as follows
\[d\sigma=d\bar{t}+F_{\sigma}(r)dr\]
\[G^{;}\equiv G(\sigma)^{;}=\frac{dG}{d\sigma}.\]
We obtain then
\begin{equation}
G^{;}F_{\sigma}(r)=\tan 2\alpha-\frac{k_{2}}{\cos 2\alpha}, k_{2}=\pm 1,
\end{equation}
so that metric (2.9) becomes
\begin{equation}
ds^{2}=-d\bar{t}^{2}-\frac{2k_{1}k_{2}}{\cos^{\frac{1}{2}}2\alpha}d\bar{t}dr+dz^{2}+r^{2}d\phi^{2},
\end{equation}
where $k_{1}$ is defined in (2.13).

Metric (3.4) still is not the kink
metric in standard form. This is obtained by making the re-definition
\begin{equation}
d\hat{t}=k_{1}\frac{d\bar{t}}{\cos^{\frac{1}{2}}2\alpha}=dt+(\tan
2\alpha-\frac{k_{2}}{\cos 2\alpha})dr,
\end{equation}
so that we finally obtain
\begin{equation}
ds^{2}=-\cos 2\alpha d\hat{t}^{2}-2k_{2}d\hat{t}dr+dz^{2}+r^{2}d\phi^{2}.
\end{equation}
Metric (3.6) has the same form
as the general line element in standard form given by
Finkelstein and McCollum [16], after replacing spherical symmetry for
cylindric symmetry, and becomes the solution of the same
Einstein equation as for (2.1) when this equation is expressed in
terms of the new time coordinate $\hat{t}$.
On the other hand, using (2.10) it can be
readily seen that any $z$=const section of metric (3.6) is not
but the hemispherical
section of the standard De Sitter
kink metric [17], for a positive cosmological constant
$\Lambda=\frac{3}{r_{*}^{2}}$.

The choice of sign in (3.3) is
adopted for the following reason. The zeros of the denominator of
$F_{\sigma}G^{;}=(\sin2\alpha\mp 1)/\cos2\alpha$ correspond to
the two horizons where $r=r_{*}$, one per patch. For the first
patch, the horizon occurs at $\alpha=\frac{\pi}{4}$ and therefore
the upper sign is selected so that $F_{\sigma}G^{;}$ remains well
defined and hence the kink is preserved in the transformation from
(2.9) to (3.6). For the second patch the horizon occurs at
$\alpha=\frac{3\pi}{4}$ and therefore the lower sign is selected.
$k_{2}=+1$ will then correspond to the first coordinate patch,
and $k_{2}=-1$ to the second one. Thus, it is the parameter
$k_{2}$, but not $k_{1}$, which defines the two coordinate
patches needed for a complete description of the spacetime
of a cosmic string kink. One would then expect an analytical
expression for the time $\hat{t}$ entering metric (3.6) which
contains the sign ambiguity $k_{2}$, but not $k_{1}$. Taking
\begin{equation}
\sin\alpha=\frac{r}{A},
\cos\alpha=k_{2}(1-\frac{r^{2}}{2r_{*}^{2}})^{\frac{1}{2}},
\cos^{\frac{1}{2}}2\alpha=k_{1}(1-\frac{r^{2}}{r_{*}^{2}})^{\frac{1}{2}},
\end{equation}
by directly integrating (3.5)
we in fact obtain [20]
\[\hat{t}=k_{1}\int_{0/A}^{r}\frac{d\bar{t}}{\cos^{\frac{1}{2}}2\alpha}=t-r_{*}k_{2}\{\frac{A}{r_{*}}(1-\frac{r^{2}}{2r_{*}^{2}})^{\frac{1}{2}}\]
\begin{equation}
-\frac{1}{2}\ln[\frac{(A(1-\frac{r^{2}}{2r_{*}^{2}})^{\frac{1}{2}}+r_{*})(r_{*}-r)}{(A(1-\frac{r^{2}}{2r_{*}^{2}})^{\frac{1}{2}}-r_{*})(r_{*}+r)}]\},
\end{equation}
where the lower limit $0/A$ refers to the choices $r=0$ and
$r=A\equiv 2^{\frac{1}{2}}r_{*}$, depending on whether the
case $k_{2}=+1$ or the case $k_{2}=-1$ is being considered [17].
Note that all sign ambiguity arising from the square root in the
argument of the $\ln$ has been omitted in Eqn. (3.8). Such an
ambiguity, so as some constant term coming from the lower
integration limit, does not affect the discussion to follow as it
only manifests as an additive constant term which
leaves metric (3.6) unchanged. This ambiguity will be of decisive
importance however for the consideration of the thermal properties
of the extreme string which we consider in section 4.

The embedded geometry of a $t$=const, $z$=const section of an extreme
string kink is that of two hemispheres, each corresponding
to a coordinate patch, whose mutual
matching at their equators lies on the extremum imaginary
values of the spherical coordinate $\theta$,
$\ln(2^{\frac{1}{2}}+1)\rightarrow\ln(2^{\frac{1}{2}}-1)$. The
imaginary values of $\theta$ are mapped onto
real values of $\alpha$ in the kink, giving rise to an equator
identification at $r=A$ which is ensured by continuity of
the tilt angle $\alpha$ at $\alpha=\frac{\pi}{2}$. Thus, the
lightcone configuration that corresponds to the one-kink
cosmic string is that given in Fig. 1.

We finally note that, as mentioned in section 2, time $\tau$
must become generally complex for $r>r_{*}$,
\begin{equation}
d\tau=(1-\frac{r^{2}}{r_{*}^{2}})^{\frac{1}{2}}dt
+2k_{2}\frac{r}{A}\left(\frac{1-\frac{r^{2}}{2r_{*}^{2}}}{1-\frac{r^{2}}{r_{*}^{2}}}\right)^{\frac{1}{2}}dr,
\end{equation}
so that by enforcing the system to lie parallel to the $r$
axis, that is taking time $t=t_{0}$ constant, and using (3.7), we have
\[\tau=t_{0}-\frac{k_{2}r_{*}^{2}}{A}\{[(1-\frac{r^{2}}{2r_{*}^{2}})(1-\frac{r^{2}}{r_{*}^{2}})]^{\frac{1}{2}}\]
\begin{equation}
+k_{1}\frac{r_{*}}{A}\ln[\frac{2k_{1}A}{r_{*}}(1-\frac{r^{2}}{r_{*}^{2}})^{\frac{1}{2}}+4(1-\frac{r^{2}}{2r_{*}^{2}})^{\frac{1}{2}}]\}\mid_{0/A}^{r}.
\end{equation}
We can check that in fact for real $t_{0}$,
both $d\tau$ and $\tau$ are real for
$r\leq r_{*}$ and complex for $A\geq r>r_{*}$.

\section{\bf Kruskal Extension of Extreme String Metric}
\subsection{Removal of Geodesic Incompleteness}
\setcounter{equation}{0}

The geodesic incompleteness at $r=r_{*}$, which occurs in each of the
two
coordinate patches described by metric (3.6), can be
removed by the usual Kruskal technique [19]. Thus, we
define the metric
\begin{equation}
ds^{2}=-2F(U,V)dUdV+dz^{2}+r^{2}d\phi^{2},
\end{equation}
in this way straightening the null geodesics into lines parallel
to the new $U$ and $V$ axis, and identify it with the standard
metric (3.6), with $g_{\hat{t}\hat{t}}=-\cos 2\alpha$,
$g_{\hat{t}r}=-k_{2}$ and $g_{UV}=-F$, in such a way
that
$F$ be finite, nonzero and depend on $r$ and $k_{2}$ alone.
All of these requirements can be met by the choice
\begin{equation}
U=\mp e^{\beta \hat{t}}\exp(2\beta k_{2}\int_{0/A}^{r}\frac{dr}{\cos 2\alpha})
\end{equation}
\begin{equation}
V=\mp\frac{1}{\beta r_{*}}e^{-\beta \hat{t}}
\end{equation}
\begin{equation}
F=-\frac{r_{*}\cos 2\alpha}{2\beta}\exp(-2\beta
k_{2}\int_{0/A}^{r}\frac{dr}{\cos 2\alpha}),
\end{equation}
with $\beta$ a constant which should be chosen so that $F$ has a finite
limit as $r\rightarrow r_{*}$.

Using [17]
\[\int_{0/A}^{r}\frac{dr}{\cos
2\alpha}=\frac{1}{2}r_{*}\ln(\frac{r_{*}+r}{r_{*}-r}),\]
where we have omitted a constant term coming from the lower
integration limit because it is canceled by the similar constant
term which appears in the Kruskal coordinate $U$, and
\[\cos 2\alpha=1-\frac{r^{2}}{r_{*}^{2}},\]
we obtain from (4.4)
\[F=-\frac{(r_{*}^{2}-r^{2})}{2\beta
r_{*}}[\frac{(r_{*}+r)^{2}}{r_{*}^{2}-r^{2}}]^{-\beta k_{2}r_{*}}.\]

To avoid $F$ being either 0 or $\infty$ at $r=r_{*}$, we then choose
$\beta=-\frac{1}{k_{2}r_{*}}$, and arrive therefore at
\begin{equation}
F=\frac{1}{2}k_{2}(r_{*}+r)^{2}
\end{equation}
\begin{equation}
U=\mp e^{-\frac{k_{2}\hat{t}}{r_{*}}}(\frac{r_{*}-r}{r_{*}+r})
\end{equation}
\begin{equation}
V=\pm k_{2}e^{\frac{k_{2}\hat{t}}{r_{*}}} ,
\end{equation}
so that
\begin{equation}
UV=-k_{2}\frac{(r_{*}-r)}{(r_{*}+r)}.
\end{equation}

Since $\hat{t}$ depends on $k_{2}$, but not $k_{1}$, it follows that
$F$, $U$ and $V$ contain $k_{2}$, but not $k_{1}$, as well.

In terms of the coordinate product (4.8) we have finally
\begin{equation}
F=\frac{2r_{*}^{2}k_{2}}{(k_{2}-UV)^{2}}
\end{equation}
\begin{equation}
r=r_{*}(\frac{k_{2}+UV}{k_{2}-UV})
\end{equation}
\[\hat{t}=t-k_{2}r_{*}\{(1-\frac{4k_{2}UV}{(k_{2}-UV)^{2}})^{\frac{1}{2}}\]
\begin{equation}
+\frac{1}{2}\ln[\frac{(1+(1-\frac{4k_{2}UV}{(k_{2}-UV)^{2}})^{\frac{1}{2}})k_{2}UV}{1-(1-\frac{4k_{2}UV}{(k_{2}-UV)^{2}})^{\frac{1}{2}}}]\}.
\end{equation}
The metric (4.1) becomes then
\begin{equation}
ds^{2}=-\frac{4k_{2}r_{*}^{2}}{(k_{2}-UV)^{2}}dUdV+dz^{2}+\frac{r_{*}^{2}(k_{2}+UV)^{2}}{(k_{2}-UV)^{2}}d\phi^{2},
\end{equation}
which, as expected, does not contain any sign parameter other than
$k_{2}$.

We notice that the $z$=const sections of this metric actually
coincide with that is obtained for the hemispherical section
of a De Sitter kink with
positive cosmological constant $\Lambda=\frac{3}{r_{*}^{2}}$ [17].
In Fig. 2 we give a representation in terms of coordinate $U$,
$V$ of the two different
coordinate patches that occur in the
one-kink geodesically-complete extreme string spacetime. These
Kruskal diagrams do not show any spatial infinities and can only
be extended beyond $r=r_{*}$ up to the surfaces $r=A$, both on
the original and new regions created by the Kruskal extension.
Because of the continuity of the angle of tilt $\alpha$ on
$\alpha=\frac{\pi}{2}$, such surfaces should be identified in
passing from one patch to another, either on the original
regions $I$ and $II$ or on the new regions $III$ and $IV$.

The maximally-extended extreme string metric (4.12) does not
possess any singularity and covers all the space of an extreme
string kink, including both an unbroken-phase interior of radius
$r_{*}$ and a broken-phase covering shell of width
$(2^{\frac{1}{2}}-1)r_{*}$.
The first of these facts is in
sharp contrast with the unavoidability of an unwellcome
singularity found by other authors for supermassive cosmic
strings [8,9]. The fact that (4.12) also describes an
exterior shell that protects the core
from dissolving is, in turn,
in contrast with the trivial maximal extension (2.7) of the
interior metric of cosmic strings.

In order to analyse possible physical processes taking
place in the spacetime of a cosmic string kink, it is
illustrative to consider the paths followed by null
geodesics on the Kruskal diagrams. For the solid geodesic
on Fig. 2, the segment in the original region $I_{+}$
starts at the pole $r=0$,
crosses the $V$ axis on the event horizon $r=r_{*}$
and passes into the original region $II_{+}$; it then continues
along that region to pass over into the original region
$I_{-}$ of the second patch at the surface $r=A$ on which
the two coordinate patches are identified. The null geodesic
crosses finally the $U$ axis at $t=\infty$ to get in the new
region $III_{-}$, propagating along it to die at the pole
$r=0$. Since identification of surfaces at $r=0$ of the
two patches is disallowed, null geodesics can never complete
a closed itinerary.

\subsection{Physical Consequences}

The existence of an extreme string kink
which on each $z$=const section exactly possesses the
symmetry of a hemispherical section of the
De Sitter kink,
may have consequences of interest. First of all, in order for
the event horizon of the De Sitter kink at $r=r_{*}$ to be
a cosmological horizon [20] with size
$H_{0}^{-1}=(\frac{3}{8\pi GV_{0}})^{\frac{1}{2}}$,
according to (2.2), one
should take $\epsilon=\frac{V_{0}}{3}$ (with $\epsilon$
the uniform string density and $V_{0}$ the vacuum energy
of the model), so that the cosmological constant becomes
$\Lambda=\frac{3}{r_{*}^{2}}=8\pi GV_{0}$. Then the
radius of a spherical extreme string, $\sqrt{2}r_{*}$,
would exceed the size of its corresponding cosmological
horizon, $H_{0}^{-1}=r_{*}$, by precisely the width
$r_{*}(\sqrt{2}-1)$ of the true-vacuum shell of the string,
and therefore the extreme cosmic string kink can quite
naturally drive a De Sitter inflationary process [21],
without any fine tuning of the initial conditions [4,5]
or the kind of uncertainties pointed out in the Introduction.
On the other hand, since we have now $r_{*}\sim\eta/\sqrt{V_{0}}$,
and hence a symmetry-breaking scale $\eta\sim\sqrt{\mu}$
of the order the Planck scale, the
inflationary expansion driven in the extreme
string kink should be
primordial (i.e. taking place in the Planck era)
and essentially unique. It is worth noting, moreover,
that if we would interpret our one-kink cosmic strings
as incipient baby universes, then
the above implication would lead to quite
confortably accommodate the concept of an eternal process
of continually self-reproducing inflating universes [4,13]
in the present picture.

On the other hand,
identification between surfaces in $III_{k_{2}}$
and $II_{k_{2}}$ (or between surfaces in $IV_{k_{2}}$
and $I_{k_{2}}$) will simply come from explicitely displaying
the sign ambiguity of the square root of the argument for the $\ln$
in Eqn. (3.10). Thus, if we want to express that argument as an
absolute value, then one has as the most general expression for metrical
time
\[\hat{t}(k_{3})=t-r_{*}k_{2}\{\frac{A}{r_{*}}(1-\frac{r^{2}}{2r_{*}^{2}})^{\frac{1}{2}}\]
\[-\ln\{\mid[\frac{(A(1-\frac{r^{2}}{2r_{*}^{2}})^{\frac{1}{2}}+r_{*})(r_{*}-r)}{(A(1-\frac{r^{2}}{2r_{*}^{2}})^{\frac{1}{2}}-r_{*})(r_{*}+r)}]^{\frac{1}{2}}\mid\}\}+\frac{i}{2}k_{2}k_{3}(1-k_{3})\pi r_{*}\]
\begin{equation}
=\hat{t}+\frac{i}{2}k_{2}k_{3}(1-k_{3})\pi r_{*},
\end{equation}
where $\hat{t}\equiv\hat{t}(k_{2})$ is the same as in (3.10),
and $k_{3}=\pm 1$ is a new sign parameter which unfolds the two
coordinate patches given in Fig. 2 into still two sets of two
patches. Time (4.13) is the most general expression for the time
$\hat{t}$ entering the standard metric (3.6). One would again
recover metric (3.6) from metric (4.1) with the same requirements
as before using $\hat{t}(k_{3})$ instead of $\hat{t}$
if we re-define the Kruskal coordinates $U$, $V$ such that
\begin{equation}
U=\pm k_{3}e^{-\frac{k_{2}\hat{t}_{c}}{r_{*}}}\frac{(r_{*}-r)}{(r_{*}+r)}
\end{equation}
\begin{equation}
V=\mp k_{2}k_{3}e^{\frac{k_{2}\hat{t}_{c}}{r_{*}}},
\end{equation}
where
\begin{equation}
\hat{t}_{c}=\hat{t}+ik_{2}k_{3}\pi r_{*}.
\end{equation}
This choice leaves expressions (4.8), (4.9), (4.10) and, of course,
the Kruskal metric (4.12) real and unchanged.

For $k_{3}=-1$, (4.14) and (4.15) become the sign-reversed to (4.6)
and (4.7); i.e. the points $(\hat{t}-ik_{2}\pi r_{*},r,z,\phi)$
on the coordinate patches of Fig. 2 are the points on the new
region $III_{k_{2}}$, on the same figure, obtained by
reflecting in the origins of the respective $U$,$V$ planes,
while keeping metric (4.12) and the physical time $t$ real
and unchanged.
This periodicity in the metric
should result [22] in the appearance of a thermal bath of
particles which, in the present case,
entails no unjustified
extension of the physical time $t$ into the imaginary axis.
Our approach offers therefore a physically reasonable
resolution of the paradox posed by the Euclidean gravity
method [23].

Let us consider then the evolution of a field
along any null geodesics on Fig. 2 described
by a quantum propagator.
If the field has mass $m$, such
a propagator will be the same as the propagator $G(x',x)$
used by Gibbons and
Hawking [20], and satisfy therefore the Klein-Gordon equation
\begin{equation}
(\Box_{x}^{2}-m^{2})G(x',x)=-\delta(x,x').
\end{equation}
For metric (4.12), the propagator $G(x',x)$ becomes analytic [20]
on precisely the strip of width $\pi r_{*}$ predicted by (4.16),
here without any need to extend $t$ to the Euclidean regime.

{}From (4.16), we can write for the amplitude for detection of
a detector sensitive to particles of energy $E$ in regions
$II_{k_{2}}$ [20,22]
\begin{equation}
\Pi_{E}=e^{k_{2}k_{3}\pi
r_{*}E}\int_{-\infty}^{+\infty}d\hat{t}e^{-iE\hat{t}}G(0,\vec{R'};\hat{t}+ik_{2}k_{3}\pi r_{*},\vec{R}),
\end{equation}
where $\vec{R'}$ and $\vec{R}$ denote respectively $(r',z',\phi ')$
and $(r,z,\phi )$. Note that the time parameter entering the
amplitude for detection is $\hat{t}_{c}$, rather than
$t$, as both the matter field and the detector must evolve in
the spacetime described by metrics (3.6) and (4.12). Since time
$\hat{t}_{c}$ (but not $t$) already contains the imaginary term
which is exactly required for the thermal effect to appear, we have
not need to make the physical time $t$ complex.
Following Gibbons and Hawking [20], we now investigate the different
particle creation processes that can take place on the extreme
hyperbolae at $r=0$ and $r=2^{\frac{1}{2}}r_{*}$.
Let us first
consider the case $k_{3}=-1$ for the original regions
$II_{k_{2}}$ on the patches of Fig. 2. For $k_{2}=+1$,
if $x'$ is a fixed point on the hyperbola at $r=0$ of region
$I_{+}$, and $x$ is a point on the hyperbola at
$r=2^{\frac{1}{2}}r_{*}$ of region $II_{+}$, we obtain
from (4.18)
\begin{equation}
P_{a}^{II_{+}}(E)=e^{-2\pi r_{*}E}P_{e}^{II_{+}}(E),
\end{equation}
where $P_{a}^{II_{k_{2}}}(E)$ generically denotes the
probability for detector to absorb a particle with positive
energy $E$ from region $II_{k_{2}}$, and
$P_{e}^{II_{k_{2}}}(E)$ accounts for the similar
probability for detector to emit the same energy also to region
$II_{k_{2}}$ [20,22].

An observer on the extreme hyperbola of the exterior original
region of patch $k_{2}=+1$ will then measure an isotropic
background of thermal positive-energy radiation at a temperature
\begin{equation}
T_{s}=\frac{1}{2\pi r_{*}}.
\end{equation}

If, in turn, $x'$ and $x$ are fixed points on the extreme
hyperbolae in regions $I_{-}$ and $II_{-}$, respectively,
we obtain then for an observer on the hyperbola $r=0$ in the
interior original region
\begin{equation}
P_{a}^{II_{-}}(-E)=e^{2\pi r_{*}E}P_{e}^{II_{-}}(-E).
\end{equation}
According to (4.21), there will appear an isotropic background
of thermal radiation which is formed by exactly the antiparticles
to the particles contained in the thermal bath detected in
region $II_{+}$, at the same temperature $T_{s}$ given by
(4.20). Thus, we can regard the joint process as the propagation
of thermal particles with positive energy from the second
to the first patch through just original regions.

For $k_{3}=+1$ we obtain similar hypersurface identifications
as for $k_{3}=-1$. In this case, the identification comes
about by simply
exchanging the mutual positions of the original regions
$I_{k_{2}}$ and $II_{k_{2}}$ for,
respectively, the new regions $III_{k_{2}}$ and
$IV_{k_{2}}$, on the coordinate patches given in
Fig. 2, while keeping the sign of coordinates $U$, $V$
unchanged with respect to those in (4.6) and (4.7); i.e.
the points $(\hat{t}+ik_{2}\pi r_{*},r,z,\phi)$ on
the so-modified regions are the points on the original
regions $II_{k_{2}}$, on the same patches, again
obtained by reflecting in the origins of the respective
$U$,$V$ planes, while keeping metric (4.12) and the
physical time $t$ real and unchanged.

Expressions for the relation between probabilities of absorption
and emission for $k_{3}=+1$ are obtained by simply replacing
regions $II_{k_{2}}$ for regions $III_{k_{2}}$
(or regions $I_{k_{2}}$ for regions $IV_{k_{2}}$),
and energy $E$ for $-E$, in (4.19) and (4.21), keeping the
same radiation temperature (4.20) in all the cases. Thus,
we obtain that observers on extreme hyperbolae will detect
an isotropic thermal bath of particles with energy: (i) $E<0$
in the exterior new region $III_{+}$,
and (ii) $E>0$ in the
interior new region $III_{-}$, in both cases at an equilibrium
temperature (4.20). The joint process would now represent
propagation of a thermal bath of particles with positive
energy from the first to the second patch just through the
new regions created by Kruskal extension.

\vspace{0.5cm}

\noindent {\bf Acknowledgement}

The author thanks S.W. Hawking for hospitality in
the Department of Applied Mathematics
and Theoretical Physics, University of Cambridge, UK, where part
of this work was done, and G.A.
Mena Marug\'an, of IMAFF, CSIC, Madrid, for very enlightening
discussions. This work has been supported by a CAICYT Research
Project N§ PB91-0052.

\pagebreak

\noindent\section*{References}
\begin{description}
\item [1] T.W.B. Kibble, J. Phys. A9, 1387 (1976); A. Vilenkin,
Phys. Rep. 121, 236 (1985).
\item [2] A. Vilenkin, in {300 Years of Gravitation}, eds S.W. Hawking
and W. Israel (Cambridge Univ. Press, Cambridge, 1987).
\item [3] A. Vilenkin, Phys. Rev. Lett. 46, 1169 (1981);
Phys. Rev. D24, 2082 (1981); T.W.B. Kibble and N. Turok,
Phys. Lett. 116B, 141 (1982).
\item [4] A.D. Linde and D.A. Linde, Phys. Rev. D50, 2456 (1994).
\item [5] A. Vilenkin, {\it Topological Inflation}, gr-qc/940240,
preprint (1994).
\item [6] A. Vilenkin and E.P.S. Shellard, {\it Cosmic Strings
and Other Topological Defects} (Cambridge Univ. Press, Cambridge, 1994).
\item [7] A. Vilenkin, Phys. Rev. D23, 852 (1981).
\item [8] P. Laguna and D. Garfinkle, Phys. Rev. D40, 1011 (1989).
\item [9] M.E. Ortiz, Phys. Rev. D43, 2521 (1991).
\item [10] W.A. Hiscock, Phys. Rev. D31, 3288 (1985).
\item [11] J.R. Gott, Astrophys. J. 288, 422 (1985).
\item [12] D. Finkelstein and C.W. Misner, Ann. Phys. (N.Y.) 6, 230 (1959).
\item [13] A.D. Linde, {\it Inflation and Quantum Cosmology} (Academic
Press, Boston, 1990).
\item [14] D. Kramer, H. Stephani, M. MacCallum and E. Herlt,
{\it Exact Solutions of Einstein's Field Equations} (Cambridge
Univ. Press, Cambridge, 1980).
\item [15] H.B. Nielsen and P. Olesen, Nucl. Phys. B61, 45 (1973).
\item [16] D. Finkelstein and G. McCollum, J. Math. Phys. 16, 2250 (1975).
\item [17] K.A. Dunn, T.A. Harriott and J.G. Williams, J. Math. Phys.
35, 4145 (1994).
\item [18] Note that the first term of the rhs in Eqn. (7) of
Ref. [17] contains the sign ambiguity here denoted as $k_{2}$.
This is incorrect and possibly due to a typing error.
\item [19] M.D. Kruskal, Phys. Rev. 119, 1743 (1981).
\item [20] G.W. Gibbons and S.W. Hawking, Phys. Rev. D15, 2738 (1977).
\item [21] A.H. Guth, Phys. Rev. D23, 347 (1981);
S.W. Hawking, in: {\it The Very Early Universe},
eds. G.W. Gibbons, S.W. Hawking and S.T.C. Siklos (Cambridge Univ. Press,
Cambridge, 1983).
\item [22] J.B. Hartle and S.W. Hawking, Phys. Rev. D31, 2188 (1976).
\item [23] N. Pauchapakesan, in {\it Hightlights in Gravitation
and Cosmology}, eds B.R. Iyer, A. Kembhavi, J.V. Narlikar and
C.V. Vishveshwara (Cambridge Univ. Press, Cambridge, 1988).

\end{description}

\pagebreak

\noindent {\bf Legends for Figures}.

\vspace{1cm}

\noindent $\bullet$ Fig. 1. One-kink lightcone configuration
for the extreme string kink. Also represented are geodesics passing
through the different regions $I_{k_{2}}$ and
$II_{k_{2}}$.

\vspace{.5cm}

\noindent $\bullet$ Fig. 2. The two coordinate patches for the
one-kink extended extreme string. In the figure,
$A=2^{\frac{1}{2}}r_{*}$. Each point on the diagrams represents
an infinite cylinder. The null geodesics starting at $North_{+}$
(solid line) and at $South_{-}$ (broken line) are not closed
as any pole identification is disallowed. The
curved lines at $t=0$ and all similar $t$= const. curves
can neither close up and, since their slope $dU/dV$ changes
sign at the horizons, they are not spacelike everywhere.
Thus, the extended spacetime of the extreme cosmic string
kink do not admit a complete foliation.

\end{document}